\documentclass[a4paper,11pt,dvipdfmx]{article}
\usepackage{epic,amssymb,amsmath}
\usepackage{color}
\usepackage{subfig}
\usepackage{graphicx}
\usepackage[all]{xy}
\usepackage{xypic}
\newcommand{\qed}{\hfill $\Box$}
\def\C{{\mathbb C}}
\def\Z{{\mathbb Z}}
\def\N{{\mathbb N}}
\def\P{{\mathbb P}}
\def\T{{\mathbb T}}
\def\Q{{\mathbb Q}}

\def\bx{\text{\mathversion{bold}{$x$}}}
\def\by{\text{\mathversion{bold}{$y$}}}
\def\bp{\mbox{\boldmath $p$}}
\def\DIS{\displaystyle}
\def\nn{\nonumber}
\newcommand{\bhline}[1]{\noalign{\hrule height #1}}

\newtheorem{example}{Example}
\newtheorem{proposition}{Proposition}
\newtheorem{remark}{Remark}

\begin{document}
\title{\textbf{A family of integrable and non-integrable difference equations arising from cluster algebras}}
\author{
Atsushi Nobe\\
Faculty of Education, Chiba University\footnote{1-33 Yayoi-cho, Inage, Chiba 263-8522, Japan}\\
nobe@faculty.chiba-u.jp\\
\\
Junta Matsukidaira\\
Department of Applied Mathematics and Informatics, Ryukoku University\footnote{1-5 Yokotani, Seta Oe-cho, Otsu, Shiga 520-2194, Japan}\\
junta@rins.ryukoku.ac.jp
}

\date{}

\maketitle

\begin{abstract}
The one-parameter family of second order nonlinear difference equations each of which is given by
\begin{align*}
x_{n-1}x_nx_{n+1}=x_{n-1}+(x_n)^{\beta-1}+x_{n+1}
\qquad(\beta\in\N)
\end{align*}
is explored.
Since the equation above is arising from seed mutations of a rank 2 cluster algebra, its solution is periodic only when $\beta\leq3$.
In order to evaluate the dynamics with $\beta\geq4$, 
%%%%%%%%%% revision %%%%%%%%%%
{the} % inserted
%%%%%%%%%% revision %%%%%%%%%%
algebraic entropy of the birational map equivalent to the difference equation is investigated; it vanishes when $\beta=4$ but is positive when $\beta\geq5$.
This fact suggests that the difference equation with $\beta\leq4$ is integrable but that with $\beta\geq5$ is not.
It is moreover shown that  the difference equation with $\beta\geq4$ fails the singularity confinement test.
This fact is consistent with linearizability of the equation with $\beta=4$ and reinforces non-integrability of the equation with $\beta\geq5$.
\end{abstract}

%-------------------------------------%
%-------------- SECTION --------------%
%-------------------------------------%
\section{Introduction}\label{sec:intro}
Since the introduction of cluster algebras by Fomin and Zelevinsky in the series of papers \cite{FZ02, FZ03, BFZ05, FZ06} there has been increasing interest in the studies of the Laurent phenomenon and positivity which arise in cluster algebras and characterize them extremely.
The Laurent phenomenon of cluster algebras, which states that any cluster algebra is the Laurent polynomial ring generated by its initial cluster variables, is established by Fomin and Zelevinsky in their first paper \cite{FZ02}.
It is now known that the Laurent phenomenon plays crucial roles in the studies of cluster algebras and their applications to various mathematics such as Poisson geometry, Teichm\"uller theory, mirror symmetry, dynamical systems and so forth \cite{FZ03-2, CK05, FG05, FR05, FZ06, IIKNS10}.
Moreover, generalizations of cluster algebras from the viewpoint of the Laurent phenomenon are also investigated \cite{LP12, LP16}.
On the other hand, positivity of cluster algebras, which states that any cluster variable in a cluster algebra is a subtraction-free Laurent polynomial in the initial cluster variables, is also conjectured by Fomin and Zelevinsky in \cite{FZ02}.
It has been open for a decade but finally established by Lee and Schiffler in 2013 \cite{LS13} for skew-symmetric cases and by Gross, Hacking, Keel and Kontsevich in 2014 \cite{GHKK14} for cluster algebras of geometric type.
Positivity of cluster algebras strongly promotes application of cluster algebras to combinatorial representation theory, tropical geometry and discrete integrable systems \cite{FZ03-2, SW05, FM11,Okubo13, FH14, GHKK14, Nobe16, Nakata17, HLK19}.

In this paper, we focus our attention on the dynamics of seed mutations of cluster algebras, \textit{i.e.}, we reduce a family of difference equations from the seed mutations and investigate their integrable or non-integrable structures.
More precisely, we study a one-parameter, $\beta\in\N$, family of second order nonlinear difference equations arising from seed mutations of one-parameter family of rank 2 cluster algebras.
Since the difference equations we consider are arising from cluster algebras, they inherit the Laurent phenomenon and the positivity.
Due to these properties, criteria for (non-) integrability such as the algebraic entropy \cite{HV98, BV99} and the singularity confinement \cite{GRP91} can concretely be computed via initial value problems of linear difference equations.

Every rank 2 cluster algebra is referred to as the type of its Cartan counterpart since the counterpart is uniquely determined.
Hence, the reduced difference equations can also be referred to as their types.
For the finite types, $A_2$ ($\beta=1$), $B_2$ ($\beta=2$) and $G_2$ ($\beta=3$), the difference equations are finitely periodic, \textit{i.e.}, their solutions have certain fixed periods for any initial values.
For the affine type, $A^{(2)}_2$ ($\beta=4$), the difference equation is finitely periodic no longer; nevertheless, it is still integrable.
Actually, in \cite{Nobe18}, we constructed the invariant curve and obtained the general solution by using the conserved quantity.
This gives the subtraction-free Laurent polynomial expression of the cluster variables exactly.
On the other hand, for the strictly hyperbolic type, $\beta\geq5$, the difference equation is integrable no longer.
This fact is shown by computing the algebraic entropy of the birational map equivalent to the difference equation through the recursion relation of the homogeneous degree of the map.
It should be noted that, in general, to compute the algebraic entropy of a birational map is difficult since the recursion relation does not have a simple expression (see \cite{HV98,BV99, KMT15}); whereas our recursion relation turns into a second order linear difference equation, and we have only to solve an initial value problem of it.
Positive algebraic entropy of the difference equation suggests its non-integrability.

We also show that the birational map with $\beta\geq4$ never passes the singularity confinement test. 
Remark that it is true not with $\beta\geq5$ but with $\beta\geq4$.
Failure of the test when $\beta=4$ is consistent with the fact that the difference equation is linearizable \cite{Nobe18}.
If $\beta\geq5$ we easily see numerical chaos in the iterates of the birational map, therefore, the difference equation is considered to be non-integrable, rather chaotic.
We note that such property that the dynamics of a one-parameter family of second order difference equations varies from integrable to non-integrable depending on the parameter is similar to the extended Hietarinta -- Viallet equation introduced by Kanki, Mase and Tokihiro \cite{HV98,KMT15}.

This paper is organized as follows.
In section \ref{subsec:CA} and \ref{subsec:BRM}, we briefly review cluster algebras, and reduce a one-parameter family of birational maps from a one-parameter family of rank 2 cluster algebras.
We also discuss the variable transformation which leads the birational maps to the nonlinear difference equations and the conserved quantity of them in section \ref{subsec:VT} and \ref{subsec:CQ}.
In section \ref{subsec:SCT}, we apply the singularity confinement test to the birational map, and show that it fails the test for any $\beta\geq4$.
In section \ref{subsec:AE}, we give the homogeneous degree of the iterates of the birational map by solving an initial value problem of a second order linear difference equation, and compute the algebraic entropy explicitly.
We discuss similarity of our difference equation with the extended Hietarinta -- Viallet equation in section \ref{subsec:CCA}.
Section \ref{sec:CR} is devoted to concluding remarks.

%-------------------------------------%
%-------------- SECTION --------------%
%-------------------------------------%
\section{From cluster algebras to difference equations via birational maps}
\label{sec:CADEBM}
%-------------------------------------%
%-------------- SUBSECTION --------------%
%-------------------------------------%
\subsection{Cluster algebras}
\label{subsec:CA}
We briefly recall cluster algebras \cite{FZ02, FZ03, BFZ05, FZ06}.
Let $\bx=(x_1,x_2,\ldots,x_n)$ be the set of generators of $\mathcal{F}=\mathbb{QP}(\bx)$, where $\P=\left(\P,\cdot,\oplus\right)$ is a semifield endowed with multiplication $\cdot$ and auxiliary addition $\oplus$ and $\mathbb{QP}$ is the group ring of $\P$ over $\Q$.
Also let $\by=(y_1,y_2,\ldots,y_n)$ be an $n$-tuple in $\P^n$ and $B=(b_{ij})$ be an $n\times n$ skew-symmetrizable integral matrix.
The triple $(\bx,\by,B)$ is referred to as the seed.
We also refer to $\bx$ as the cluster of the seed, to $\by$ as the coefficient tuple and to $B$ as the exchange matrix.
Elements of $\bx$ and $\by$ are respectively called the cluster variables and the coefficients.

Introduce seed mutations.
Let $k$ be a natural number equally less than $n$.
The seed mutation $\mu_k$ in the direction $k$ transforms a seed $(\bx,\by,B)$ into the seed $(\bx^\prime,\by^\prime,B^\prime):=\mu_k(\bx,\by,B)$ defined by the birational equations (\ref{eq:mutem}-\ref{eq:mutcv}) called the exchange relations:
\begin{align}
b_{ij}^\prime
&=
\begin{cases}
-b_{ij}&\mbox{$i=k$ or $j=k$},\\
b_{ij}+[-b_{ik}]_+b_{kj}+b_{ik}[b_{kj}]_+&\mbox{otherwise},\\
\end{cases}
\label{eq:mutem}\\
y_j^\prime
&=
\begin{cases}
y_k^{-1}&\mbox{$j=k$},\\
y_jy_k^{[b_{kj}]_+}(y_k\oplus 1)^{-b_{kj}}&\mbox{$j\neq k$},\\
\end{cases}
\label{eq:mutcoef}\\
x_j^\prime
&=
\begin{cases}
\DIS\frac{y_k\prod x_i^{[b_{ik}]_+}+\prod x_i^{[-b_{ik}]_+}}{(y_k\oplus 1)x_k}&\mbox{$j=k$},\\
x_j&\mbox{$j\neq k$},\\
\end{cases}
\label{eq:mutcv}
\end{align}
where we define $[a]_+:=\max[a,0]$ for $a\in\Z$.

Let $\T_n$ be the $n$-regular tree.
The edges in $\T_n$ are labeled by $1, 2, \ldots, n$ so that the $n$ edges emanating from each vertex receive different labels.
Assign a seed $\Sigma_t=(\bx_t,\by_t,B_t)$ to every vertex $t\in\T_n$ so that the seeds assigned to the endpoints of any edge labeled by $k$ are obtained from each other by the seed mutation in direction $k$.
We refer to the assignment $\T_n\ni t\mapsto\Sigma_t$ as a cluster pattern.
Denote the elements of $\Sigma_t$ by
\begin{align*}
\bx_t=(x_{1;t},\ldots,x_{n;t}),\quad
\by_t=(y_{1;t},\ldots,y_{n;t}),\quad
B_t=(b_{ij}^t).
\end{align*}

Given a cluster pattern $\T_n\ni t\mapsto\Sigma_t$, we denote the union of clusters of all seeds in the pattern by
\begin{align*}
\mathcal{X}
=
\bigcup_{t\in\T_n}\bx_t
=
\left\{
x_{i;t}\ |\ t\in\T_n,\ 1\leq i\leq n
\right\}.
\end{align*}
The cluster algebra $\mathcal{A}$ associated with a given cluster pattern is the $\mathbb{ZP}$-subalgebra of the ambient field $\mathcal{F}$ generated by all cluster variables: $\mathcal{A}=\mathbb{ZP}[\mathcal{X}]$.
The set $\mathcal{X}$ of all cluster variables are referred to as the set of generators of $\mathcal{A}$.
The cluster algebra $\mathcal{A}$ is also generated by its initial cluster variables $\bx_0$ as a subtraction-free Laurent polynomial subring of the ambient field $\mathcal{F}$ \cite{FZ02, LS13,GHKK14}.
The number $n$ of the elements in a cluster $\bx_t$ is called the rank of $\mathcal{A}$.

%-------------------------------------%
%-------------- SUBSECTION --------------%
%-------------------------------------%
\subsection{Birational maps}
\label{subsec:BRM}
Hereafter, we fix the rank $n$ of $\mathcal{A}$ to be $2$.
Let $\beta$ be a natural number.
Consider the initial seed $\Sigma_0=(\bx_0,\by_0,B_0)$ given by
\begin{align}
\bx_0
&=\left(x_{1:0},x_{2:0}\right),
\qquad
\by_0
=\left(y_{1:0},y_{2:0}\right),
\nn\\
B_0
&=
\left(\begin{matrix}0&b_{12}^0\\ b_{21}^0&0\\\end{matrix}\right)
=
\left(\begin{matrix}0&-1\\ \beta&0\\\end{matrix}\right).
\label{eq:initialseed}
\end{align}

By applying the seed mutations $\mu_k$ in direction $k$ ($k=1,2$) alternately, we obtain the sequence of the mutated seeds:
\begin{align}
\Sigma_0
=\left(\bx_0,\by_0,B_0\right)
\
&\overset{\mu_1}{\longleftrightarrow}
\
\Sigma_1
=\left(\bx_1,\by_1,B_1\right)
\
\overset{\mu_2}{\longleftrightarrow}
\
\Sigma_2
=\left(\bx_2,\by_2,B_2\right)
\nn\\
&\overset{\mu_1}{\longleftrightarrow}
\
\Sigma_3
=\left(\bx_3,\by_3,B_3\right)
\
\overset{\mu_2}{\longleftrightarrow}
\
\Sigma_4
=\left(\bx_4,\by_4,B_4\right)
\nn\\
&\overset{\mu_1}{\longleftrightarrow}
\
\cdots.
\label{eq:seqmut}
\end{align}
These seeds are in the cluster pattern
\begin{align*}
\xymatrix{
\ar@{--}[r]^{2}
&
t_{-2}\ar@{-}[r]^{1}
&
t_{-1}\ar@{-}[r]^{2}
&
t_0\ar@{-}[r]^{1}
&
t_1\ar@{-}[r]^{2}
&
t_2\ar@{--}[r]^{1}
&
}
\end{align*}
denoted by $\T_2\ni t_m\mapsto\Sigma_m$ ($m\in\Z$).

Let us associate the variables $x^t$ and $y^t$ with the seeds $\Sigma_{2t}$ as
\begin{align}
x^t
=
\frac{x_{1;2t}}{{y_{2;2t}}},
\qquad
y^t
=
\sqrt[\beta]{y_{1;2t}}x_{2;2t}
\label{eq:variablesetting}
\end{align}
for $t\geq0$.
If we assume $(x^0,y^0)\in\P^2(\C)$ then the sequence of seed mutations \eqref{eq:seqmut} leads to the birational map $\varphi_\beta$ on $\P^2(\C)$:
\begin{align}
&\varphi_\beta:(x^t,y^t)\mapsto(x^{t+1},y^{t+1}),
\nn\\
&x^{t+1}
=
\frac{\left(y^t\right)^\beta+1}{x^t},
\qquad
y^{t+1}
=
\frac{x^{t+1}+1}{y^t}.
\label{eq:bmap}
\end{align}

Actually, we see from (\ref{eq:mutem}) that the exchange matrix $B_m$  has period two:
\begin{align*}
B_m
=
\begin{cases}
B_0&\mbox{$m$ even,}\\
-B_0&\mbox{$m$ odd.}\\
\end{cases}
\end{align*}
From the exchange relation (\ref{eq:mutcoef}) of the coefficients, it immediately follows the equalities among them:
\begin{align*}
y_{1;2k}y_{1;2k+1}
&=
1,
\\
y_{2;2k+1}
&=
y_{2;2k}(y_{1;2k}\oplus 1),
\\
y_{2;2k+1}y_{2;2k+2}
&=
1,
\\
y_{1;2k+2}
&=
y_{1;2k+1}(y_{2;2k+1}\oplus 1)^{\beta}.
\end{align*}
It follows that we have
\begin{align*}
y_{1;2k}y_{1;2k+2}=(y_{2;2k+1}\oplus 1)^{\beta},
\qquad
y_{2;2k}y_{2;2k+2}(y_{1;2k}\oplus 1)=1.
\end{align*}

By using the exchange relation (\ref{eq:mutcv}) of the cluster variables and the above equalities in the coefficients, we compute
\begin{align*}
x_{1;2k+1}
&=
\frac{y_{1;2k}x_{2;2k}^\beta+1}{(y_{1;2k}\oplus 1)x_{1;2k}}
=
\frac{y_{1;2k}x_{2;2k}^\beta+1}{\DIS\frac{x_{1;2k}}{y_{2;2k}y_{2;2k+2}}},
\\
x_{2;2k+1}
&=
x_{2;2k},
\\
x_{1;2k+2}
&=
x_{1;2k+1}
=
\frac{y_{1;2k}x_{2;2k}^\beta+1}{\DIS\frac{x_{1;2k}}{y_{2;2k}y_{2;2k+2}}}
=
\frac{\left(\sqrt[\beta]{y_{1;2k}}x_{2;2k}\right)^\beta+1}{\DIS\frac{1}{y_{2;2k+2}}\DIS\frac{x_{1;2k}}{y_{2;2k}}},
\\
x_{2;2k+2}
&=
\frac{y_{2;2k+1}x_{1;2k+1}+1}{(y_{2;2k+1}\oplus 1)x_{2;2k+1}}
=
\frac{\DIS\frac{x_{1;2k+2}}{y_{2;2k+2}}+1}{\sqrt[\beta]{y_{1;2k+2}}\sqrt[\beta]{y_{1;2k}}x_{2;2k}}.
\end{align*}
By setting \eqref{eq:variablesetting}, we obtain \eqref{eq:bmap}.

%//////////////////// REMARK ////////////////////%
\begin{remark}%\normalfont
Since the birational map $\varphi_\beta$ is arising from the successive seed mutations $\mu_1$ and $\mu_2$, it is natural to consider $\varphi_\beta$ to be the composition $\varphi_{\beta,2}\circ\varphi_{\beta,1}$ of the following two birational maps
\begin{align*}
&\varphi_{\beta,1}:(x,y)\mapsto
\left(
\frac{y^\beta+1}{x},
y
\right),
\nn\\
&\varphi_{\beta,2}:(x,y)\mapsto
\left(
x,
\frac{x+1}{y}
\right),
\end{align*}
which are arising from the seed mutations $\mu_1$ and $\mu_2$, respectively.
The map $\varphi_{\beta,1}$ is often referred to as the horizontal flip since it moves the $x$-component only.
Also, the map $\varphi_{\beta,2}$ is often referred to as the vertical flip since it moves the $y$-component only.
This structure of the map $\varphi_\beta$ decomposing into the horizontal and the vertical flips is similar to the celebrated QRT map \cite{QRT89, Tsuda04}.
\end{remark}
%//////////////////// REMARK ////////////////////%

The Cartan counterpart $A(B_m)$ of the exchange matrix $B_m$ is given by
\begin{align}
A(B_m)
&=
\left(\begin{matrix}2&-|b_{12}^m|\\ -|b_{21}^m|&2\\\end{matrix}\right)
=
\left(\begin{matrix}2&-1\\ -\beta&2\\\end{matrix}\right)
\label{eq:initialCartan}
\end{align}
for any $m\geq0$.
Thus each cluster algebra generated from the initial seed $\Sigma_0$ is referred to as the type of the Cartan counterpart $A(B_m)$ (see table \ref{tab:Dynkin}).
The type depends on $\beta$ via the corresponding Dynkin diagram.
We also refer to the birational map $\varphi_\beta$ as the type of $A(B_m)$.

%//////////////////// TABLE ////////////////////%
\begin{table}[htbp]\centering
\caption{Classification of the birational map $\varphi_\beta$ in terms of $\beta$.
For each $\beta$, the type of the Cartan counterpart $A(B_m)$ and the Dynkin diagram are listed.
In the Dynkin diagrams, empty and filled circles stand for the vertices labeled by $1$ and by $2$, respectively.
The period of $\varphi_\beta$, the result of the singularity confinement test for $\varphi_\beta$ and the algebraic entropy of $\varphi_\beta$, which will be discussed later, are also listed.
}
\label{tab:Dynkin}
{\renewcommand\arraystretch{1.2}
\begin{tabular}{lll | lll}
\bhline{1pt}
$\beta$&Type&Dynkin diag.&Period&SC test&Alg. entropy
\\
\bhline{.2pt}
1
&
Finite: $A_2$
&
\unitlength=.06in{\def\arraystretch{1.0}
\begin{picture}(12,4)(0,-.5)
\thicklines
\put(1,0){\circle{2}}
\put(10,0){\circle*{2}}
\put(2,0){\line(1,0){7}}
\end{picture}}
&
5
&
Pass
&
$0$
\\
2
&
Finite: $B_2$
&
\unitlength=.06in{\def\arraystretch{1.0}
\begin{picture}(12,4)(0,-.5)
\thicklines
\put(1,0){\circle{2}}
\put(10,0){\circle*{2}}
\put(1.9,.5){\vector(1,0){7.2}}
\put(1.9,-.5){\vector(1,0){7.2}}
\end{picture}}
&
3
&
Pass
&
$0$
\\
3
&
Finite: $G_2$
&
\unitlength=.06in{\def\arraystretch{1.0}
\begin{picture}(12,4)(0,-.5)
\thicklines
\put(1,0){\circle{2}}
\put(10,0){\circle*{2}}
\put(1.9,.5){\vector(1,0){7.1}}
\put(2,0){\vector(1,0){7}}
\put(1.9,-.5){\vector(1,0){7.1}}
\end{picture}}
&
4
&
Pass
&
$0$
\\
4
&
Affine: $A^{(2)}_2$
&
\unitlength=.06in{\def\arraystretch{1.0}
\begin{picture}(12,4)(0,-.5)
\thicklines
\put(1,0){\circle{2}}
\put(10,0){\circle*{2}}
\put(1.8,.6){\vector(1,0){7.3}}
\put(1.9,.2){\vector(1,0){7.2}}
\put(1.9,-.2){\vector(1,0){7.2}}
\put(1.8,-.6){\vector(1,0){7.3}}
\end{picture}}
&
$\infty$
&
Fail
&
$0$
\\
$\geq5$
&
Strictly hyperbolic
&
\unitlength=.06in{\def\arraystretch{1.0}
\begin{picture}(12,4)(0,-.5)
\thicklines
\put(1,0){\circle{2}}
\put(10,0){\circle*{2}}
\put(1.8,.6){\vector(1,0){7.3}}
\put(5,-.38){$\cdot$}
\put(5,-.7){$\cdot$}
\put(5,-1.02){$\cdot$}
\put(1.8,-.7){\vector(1,0){7.3}}
\end{picture}}
&
$\infty$
&
Fail
&
$+$
\\
\bhline{1pt}
\end{tabular}
}
\end{table}
%//////////////////// TABLE ////////////////////%

The birational map $\varphi_\beta$ with $\beta\leq3$ is of finite type $A_2$, $B_2$ or $G_2$ and has period 5, 3 or 4, in order.
The period is defined to be the smallest $p>0$ such that $\varphi^{t+p}=\varphi^t$ holds for any $t\geq0$.
The map $\varphi_4$ with $\beta=4$ is of affine type $A^{(2)}_2$ and does not have a finite period; nevertheless, it is still an integrable map which has the property of linearizability \cite{Nobe18}.
Every $\varphi_\beta$ with $\beta\geq5$ is of strictly hyperbolic type and does not have a finite period, as well.
In the rest of this paper, we will show that such $\varphi_\beta$ has positive algebraic entropy and fails the singularity confinement test, both of which suggest its non-integrability.

%-------------------------------------%
%-------------- SUBSECTION --------------%
%-------------------------------------%
\subsection{Variable transformation}
\label{subsec:VT}
Let us introduce the variable transformation $\widetilde\pi$ on $\P^2(\C)$
\begin{align*}
\widetilde\pi:
(u,v)\mapsto(x,y)=(uv-1,v).
\end{align*}
Note that the inverse $\widetilde\pi^{-1}$ is uniquely determined by
\begin{align*}
\widetilde\pi^{-1}:
(x,y)\mapsto(u,v)=\left(\frac{x+1}{y},y\right)
\end{align*}
except for $(x,y)=(-1,0)$.

We obtain the conjugate $\widetilde\varphi_\beta$ of the birational map $\varphi_\beta$ with respect to $\widetilde\pi$ as follows
\begin{align}
&\widetilde\varphi_\beta:=\widetilde\pi^{-1}\circ\varphi_\beta\circ\widetilde\pi:
(u^t,v^t)\mapsto(u^{t+1},v^{t+1});
\nn\\
&u^{t+1}
=
v^t,
\qquad
v^{t+1}
=
\frac{u^t+\left(v^t\right)^{\beta-1}}{u^tv^t-1}.
\label{eq:bmapbu}
\end{align}
Eliminate $u^t$ from \eqref{eq:bmapbu}.
Then, by putting $x_n:=v^t$, we obtain the second order difference equation for $x_n$
\begin{align}
x_{n-1}x_nx_{n+1}=x_{n-1}+(x_n)^{\beta-1}+x_{n+1},
\label{eq:2dds} 
\end{align}
which coincides with the one in the abstract.

%//////////////////// REMARK ////////////////////%
\begin{remark}
The birational maps conjugate to the horizontal flip $\varphi_{\beta,1}$ and the vertical flip $\varphi_{\beta,2}$ with respect to $\widetilde\pi$ are respectively given by
\begin{align*}
&\widetilde{\varphi}_{\beta,1}:=\widetilde\pi^{-1}\circ\varphi_{\beta,1}\circ\widetilde\pi:(u,v)\mapsto
\left(
\frac{u+v^{\beta-1}}{uv-1},
v
\right),
\nn\\
&\widetilde{\varphi}_{\beta,2}:=\widetilde\pi^{-1}\circ\varphi_{\beta,2}\circ\widetilde\pi:(u,v)\mapsto
\left(
v,u
\right).
\end{align*}
Since the map $\widetilde{\varphi}_{\beta,1}$ moves the $u$-component only, it is the horizontal flip as well as the original one $\varphi_{\beta,1}$.
On the other hand, we call the map $\widetilde{\varphi}_{\beta,2}$ the diagonal flip since it exchanges the $u$ and $v$-components, while the original one $\varphi_{\beta,2}$ is the vertical flip.
Thus the map $\widetilde\varphi_\beta$ is the composition $\widetilde{\varphi}_{\beta,2}\circ\widetilde{\varphi}_{\beta,1}$ of the horizontal flip $\widetilde{\varphi}_{\beta,1}$ and the diagonal flip $\widetilde{\varphi}_{\beta,2}$.
\end{remark}
%//////////////////// REMARK ////////////////////%

A geometric meaning of the variable transformation in terms of $\widetilde\pi$ is given as follows.
Since the birational map $\varphi_4$ ($\beta=4$) is integrable, it has the quartic invariant curve $\gamma_\lambda$ defined by
\begin{align*}
f(x,y)=(x+1)^2+y^4-\lambda{xy^2},
\end{align*}
where $\lambda$ is the conserved quantity \cite{Nobe18}.
The curve $\gamma_\lambda$ has a singularity at the ordinary double point $(x,y)=(-1,0)$.
If we blow-up the curve $\gamma_\lambda$ at $(x,y)=(-1,0)$ by using $\widetilde\pi$ we obtain the non-singular quadratic curve $\widetilde\gamma_\lambda$ defined by
\begin{align*}
\widetilde{f}(u,v)=u^2+v^2-\lambda(uv-1).
\end{align*}
The curve $\widetilde\gamma_\lambda$ is of course the invariant curve of the map $\widetilde\varphi_4$.
Note that the singular point $(x,y)=(-1,0)$, at which $\widetilde\pi$ is not invertible, is also the base point of the pencil $\left\{\gamma_\lambda\right\}_{\lambda\in\P^1(\C)}$ of the curve $\gamma_\lambda$.
Therefore, we can resolve the singularity of the curves in the pencil $\left\{\gamma_\lambda\right\}_{\lambda\in\P^1(\C)}$ by the blowing-up, all at once.
Through this procedure, we find that the birational map $\varphi_4$ is linearizable and the general solution is concretely constructed by using the hyperbolic functions \cite{Nobe18}.
Note that it is difficult to find the linearizability of the map $\varphi_4$ unless we apply the blowing-up $\widetilde\pi$.
For other $\beta$, it seems that the orbits of the map $\widetilde\varphi_\beta$ are reduced simpler than the ones of original $\varphi_\beta$.
(Note that there is no invariant curve of $\widetilde\varphi_\beta$ with $\beta\geq5$.)
Throughout this paper, we choose the conjugate $\widetilde\varphi_\beta$ or the original $\varphi_\beta$ in accordance with the purpose.

%-------------------------------------%
%-------------- SUBSECTION --------------%
%-------------------------------------%
\subsection{Conserved quantities}
\label{subsec:CQ}
Let the sequence $\{x_n\}_{n\geq0}$ of points satisfying \eqref{eq:2dds} be $\mathcal{P}_\beta$.
Then, $\mathcal{P}_\beta$ with $\beta\leq3$ is a finite set. 
Actually, if $\beta=2$ we have
\begin{align*}
x_0x_1x_2&=x_0+x_1+x_2,
\\
x_1x_2x_3&=x_1+x_2+x_3.
\end{align*}
Subtracting both sides, we have
\begin{align*}
(x_0-x_3)x_1x_2=x_0-x_3.
\end{align*}
Since $x_0$ and $x_1$ are arbitrarily chosen, we have $x_3=x_0$, which implies
\begin{align*}
\mathcal{P}_2
&=
\left\{x_0,x_1,x_2\right\}.
\end{align*}
We similarly obtain
\begin{align*}
\mathcal{P}_1
&=
\left\{x_0,x_1,x_2,x_3,x_4\right\},
\\
\mathcal{P}_3
&=
\left\{x_0,x_1,x_2,x_3\right\}
\end{align*}
with $\beta=1,3$, respectively (see table \ref{tab:Dynkin}).

On the other hand, $\mathcal{P}_\beta$ with $\beta\geq4$ is an infinite set. 
This is a direct consequence of the finite type classification of cluster algebras obtained by Fomin and Zelevinsky in \cite{FZ03}.

The finiteness of the set $\mathcal{P}_\beta$ immediately implies the conserved quantity of the system.
It is easy to see that the difference equation \eqref{eq:2dds}, or the birational map $\widetilde\varphi_\beta$, with $\beta\leq3$ has the conserved quantities listed in table \ref{tab:QRTbu}.
%//////////////////// TABLE ////////////////////%
\begin{table}[htbp]\centering
\caption{
The conserved quantities of the difference equation \eqref{eq:2dds} with $\beta\leq4$.
}
\label{tab:QRTbu}
\vspace{.1cm}

{\renewcommand\arraystretch{1.2}
\begin{tabular}{lll}
\bhline{1pt}
$\beta$&Type&Conserved quantity
\\
\bhline{.2pt}
1
&
$A_2$
&
$\DIS\frac{(x_0^2x_1+1)+(x_0x_1^2+1)+(x_0x_1-1)^2}{x_0x_1-1}$
\rule[0mm]{0mm}{7mm}
\\
2
&
$B_2$
&
$\DIS\frac{(x_0+x_1)x_0x_1}{x_0x_1-1}$
\rule[0mm]{0mm}{7mm}
\\
3
&
$G_2$
&
$\DIS\frac{(x_0^2+x_1)x_1+(x_0+x_1^2)x_0}{x_0x_1-1}$
\rule[0mm]{0mm}{7mm}
\\
4
&
$A^{(2)}_2$
&
$\DIS\frac{x_0^2+x_1^2}{x_0x_1-1}$
\rule[-5mm]{0mm}{12mm}
\\
\bhline{1pt}
\end{tabular}
}
\end{table}
%//////////////////// TABLE ////////////////////%

Although the map $\widetilde\varphi_\beta$ with $\beta\geq5$ does not have the conserved quantity, the horizontal flip $\widetilde{\varphi}_{\beta,1}$ composing $\widetilde\varphi_\beta=\widetilde{\varphi}_{\beta,2}\circ\widetilde{\varphi}_{\beta,1}$ has the conserved quantity with any $\beta$.
Actually, we have
\begin{align*}
&\widetilde{\varphi}_{\beta,1}:
uv-1
{\longmapsto}
\frac{u+v^{\beta-1}}{uv-1}\times v-1
=
\frac{v^\beta+1}{uv-1}
\\
&\widetilde{\varphi}_{\beta,1}:
u^2+v^{\beta-2}
{\longmapsto}
\left(\frac{u+v^{\beta-1}}{uv-1}\right)^2+v^{\beta-2}
=
\frac{(v^\beta+1)(u^2+v^{\beta-2})}{(uv-1)^2},
\end{align*}
and hence have
\begin{align*}
\widetilde{\varphi}_{\beta,1}:
\frac{u^2+v^{\beta-2}}{uv-1}
{\longmapsto}
\frac{u^2+v^{\beta-2}}{uv-1}.
\end{align*}
Thus the map $\widetilde{\varphi}_{\beta,1}$ with any $\beta$ has the conserved quantity
\begin{align}
\frac{u^2+v^{\beta-2}}{uv-1}.
\label{eq:localcq}
\end{align}

On the other hand, the diagonal flip $\widetilde{\varphi}_{\beta,2}$ acts on \eqref{eq:localcq} as follows
\begin{align*}
\widetilde{\varphi}_{\beta,2}:
\frac{u^2+v^{\beta-2}}{uv-1}
{\longmapsto}
\frac{u^{\beta-2}+v^2}{uv-1}.
\end{align*}
The map $\widetilde{\varphi}_{\beta,2}$  fixes \eqref{eq:localcq} if and only if $\beta=4$.
Therefore, only the map $\widetilde{\varphi}_4=\widetilde{\varphi}_{4,2}\circ\widetilde{\varphi}_{4,1}$ ($\beta=4$) has the conserved quantity
\begin{align*}
\frac{u^2+v^2}{uv-1}
\end{align*}
among $\beta\geq4$ (see table \ref{tab:QRTbu}).

%-------------------------------------%
%-------------- SECTION --------------%
%-------------------------------------%
\section{Integrability and non-integrability of difference equations}
\label{sec:INI}
In this section, we assume $\beta\geq4$ in order to consider infinitely-periodic cases.
%-------------------------------------%
%-------------- SUBSECTION --------------%
%-------------------------------------%
\subsection{Singularity confinement}
\label{subsec:SCT}
Now we apply the singularity confinement test \cite{GRP91} to the birational map $\varphi_\beta$ given by \eqref{eq:bmap}.
Let us introduce the homogeneous coordinate $(x,y)\mapsto[X:Y:Z]=[x:y:1]$ on $\P^2(\C)$.
Then the map $\varphi_\beta$ reduces to the following homogeneous one:
\begin{align*}
[X^t:Y^t:Z^t]&\mapsto[X^{t+1}:Y^{t+1}:Z^{t+1}],
\\
X^{t+1}
&=
(Y^t)^{\beta+1}+Y^t(Z^t)^\beta,
\\
Y^{t+1}
&=
(Y^t)^\beta Z^t+(Z^t)^{\beta+1}+X^t(Z^t)^{\beta},
\\
Z^{t+1}
&=
X^tY^t(Z^t)^{\beta-1}.
\end{align*}
The homogeneous degree of the map $\varphi_\beta$ is $\beta+1$ and $\varphi_\beta$ is still subtraction-free.

To evaluate a singularity of the map $\varphi_\beta$,  we choose a special point $[X^0:Y^0:Z^0]=[0:y:1]$ as its initial value.
Then we compute
\begin{align*}
X^{1}
&=
\left(1+y^\beta\right)y,
&
&&
X^{2}
&=
(1+y^\beta)^{\beta+1},
&
&&
X^{3}
&=
0,
\\
Y^{1}
&=
1+y^\beta,
&
\longrightarrow
&
&
Y^{2}
&=
0,
&
\longrightarrow
&
&
Y^{3}
&=
0,
\\
Z^{1}
&=
0,
&
&&
Z^{2}
&=
0,
&
&&
Z^{3}
&=
0.
\end{align*}
Since the point $[X^3:Y^3:Z^3]=[0:0:0]$ is not included in $\P^2(\C)$, we conclude that the birational map $\varphi_\beta$ has a singularity at the point $[0:y:1]$.

Let us apply the singularity confinement test to the map $\varphi_\beta$ at the singular point $[0:y:1]$.
For $\varepsilon>0$, we choose the initial value $[X^0:Y^0:Z^0]=[\varepsilon:y:1]$ in stead of the singular point $[0:y:1]$.
We then have
\begin{align*}
X^{1}
&=
\left(1+y^\beta\right)y,
\\
Y^{1}
&=
1+y^\beta+\varepsilon,
\\
Z^{1}
&=
y\varepsilon
\end{align*}
for $t=1$ and
\begin{align*}
X^{2}
&=
(1+y^\beta)^{\beta+1}+(\beta+1)(1+y^\beta)^{\beta}\varepsilon+o(\varepsilon),
\\
Y^{2}
&=
(1+y^\beta)^\beta y\varepsilon+o(\varepsilon),
\\
Z^{2}
&=
\left(1+y^\beta\right)^2y^\beta\varepsilon^{\beta-1}+o(\varepsilon^{\beta-1})
\end{align*}
for $t=2$.
We further compute
\begin{align}
X^{3}
&=
(1+y^\beta)^{\beta^2+\beta} y^{\beta+1}\varepsilon^{\beta+1}+o(\varepsilon^{\beta+1}),
\nn\\
Y^{3}
&=
(1+y^\beta)^{\beta^2+2} y^{2\beta}\varepsilon^{2\beta-1}+o(\varepsilon^{2\beta-1}),
\label{eq:tequals3}\\
Z^{3}
&=
(1+y^\beta)^{4\beta-1}y^{\beta^2-\beta+1}\varepsilon^{\beta^2-2\beta+2}+o(\varepsilon^{\beta^2-2\beta+2})
\nn
\end{align}
for $t=3$.
In the homogeneous coordinate, we have
\begin{align*}
\left[X^3:Y^3:Z^3\right]
\sim
\left[
(1+y^\beta)^{\beta^2+\beta} y^{\beta+1}:
(1+y^\beta)^{\beta^2+2} y^{2\beta}\varepsilon^{\beta-2}:
(1+y^\beta)^{4\beta-1}y^{\beta^2-\beta+1}\varepsilon^{\beta^2-3\beta+1}
\right]
\end{align*}
for sufficiently small $\varepsilon>0$.
Since we assume $\beta\geq4$, we have
\begin{align*}
\left[X^3:Y^3:Z^3\right]
\to
\left[
(1+y^\beta)^{\beta^2+\beta} y^{\beta+1}:
0:
0
\right]
=
[1:0:0]
\end{align*}
in the limit $\varepsilon\to0$.
Thus the singularity has not been confined yet, which means that the information of the initial value ($y$) vanishes due to the singularity and has not been recovered yet.
We moreover show that the singularity at $[0:y:1]$ is never confined.
%//////////////////// PROPOSITION ////////////////////%
\begin{proposition}%\normalfont
Let $\varepsilon$ be a positive number.
If $[X^0:Y^0:Z^0]=[\varepsilon:y:1]$ then we have
\begin{align*}
\lim_{\varepsilon\to0}[X^t:Y^t:Z^t]
=
[1:0:0]
\end{align*}
for $t\geq3$.
\end{proposition}
%//////////////////// PROPOSITION ////////////////////%

(Proof)\quad
Let us denote the minimal degree of $\varepsilon$ in $X^t$, $Y^t$ and $Z^t$ by $d_X^t$, $d_Y^t$ and $d_Z^t$, respectively.
From \eqref{eq:tequals3}, we have
\begin{align}
&
d_X^3=\beta+1\geq5,
&&
d_Y^3=2\beta-1\geq7,
&&
d_Z^3=\beta^2-2\beta+2\geq10
\label{eq:initialdegree}
\end{align}
for $t=3$ since $\beta\geq4$.

Since the map $\varphi_\beta$ is subtraction-free, the recursion relations for $d_X^t$, $d_Y^t$ and $d_Z^t$ are given as follows
\begin{align}
d_X^{t+1}
&=
\min\left[
(\beta+1)d_Y^t,
d_Y^t+\beta d_Z^t
\right],
\nn\\
d_Y^{t+1}
&=
\min\left[
\beta d_Y^t+d_Z^t,
(\beta+1) d_Z^t,
d_X^t+\beta d_Z^t
\right],
\label{eq:setofrr}\\
d_Z^{t+1}
&=
d_X^t+d_Y^t+(\beta-1)d_Z^t.
\nn
\end{align}

Now we assume 
\begin{align}
2(d_Z^t-d_Y^t)>d_Z^t-d_X^t>d_Z^t-d_Y^t>0.
\label{eq:assumption}
\end{align}
Then the system \eqref{eq:setofrr} of recursion relations reduces to the system of linear difference equations
\begin{align}
d_X^{t+1}
&=
(\beta+1)d_Y^t,
\nn\\
d_Y^{t+1}
&=
\beta d_Y^t+d_Z^t,
\label{eq:recdegree}\\
d_Z^{t+1}
&=
d_X^t+d_Y^t+(\beta-1)d_Z^t.
\nn
\end{align}

We show that, by induction on $t$, the condition \eqref{eq:assumption} holds for any $t\geq3$ and the degrees $d_X^t$, $d_Y^t$ and $d_Z^t$ are uniquely determined by the initial value problem of the difference equation \eqref{eq:recdegree}.
We easily see that
\begin{align*}
2(d_Z^3-d_Y^3)>d_Z^3-d_X^3>d_Z^3-d_Y^3>0
\end{align*}
holds for $t=3$ (see \eqref{eq:initialdegree}).

Assume that the condition \eqref{eq:assumption} holds for $t$.
We then show that the degrees $d_X^{t+1}$, $d_Y^{t+1}$ and $d_Z^{t+1}$ for $t+1$ also satisfy \eqref{eq:assumption}.
By using \eqref{eq:recdegree}, we have
\begin{align*}
d_Z^{t+1}-d_Y^{t+1}
&=
(\beta-1)\left(d_Z^t-d_Y^t\right)-\left(d_Z^t-d_X^t\right)>0,
\\
d_Z^{t+1}-d_X^{t+1}
&=
(\beta-1)\left(d_Z^t-d_Y^t\right)-\left(d_Y^t-d_X^t\right)
>
d_Z^{t+1}-d_Y^{t+1}
>0
\end{align*}
and
\begin{align*}
2\left(d_Z^{t+1}-d_Y^{t+1}\right)-\left(d_Z^{t+1}-d_X^{t+1}\right)
&=
(\beta-2)\left(d_Z^t-d_Y^t\right)-(d_Z^t-d_X^t)>0,
\end{align*}
where we use the fact $\beta\geq4$.
Thus the degrees $d_X^{t}$, $d_Y^{t}$ and $d_Z^{t}$ for $t\geq4$ are uniquely determined by solving \eqref{eq:recdegree} with imposing the initial condition \eqref{eq:initialdegree}.

It is easy to see that we have
\begin{align*}
\lim_{\varepsilon\to0}[X^t:Y^t:Z^t]
=
[1:0:0]
\end{align*}
for $t\geq3$ since the degrees satisfy the following condition
\begin{align*}
0<d_X^t<d_Y^t<d_Z^t
\end{align*}
for $t\geq3$ reduced from \eqref{eq:assumption}.
\qed

Thus we show that the singularity at $[0:y:1]$ is never confined.
For $\beta=4$, this fact is consistent with the property that the space of initial values of a linearizable birational map can not be constructed via a finite number of blowing-ups \cite{DF01, Mase18}.
This fact also suggests that the birational map $\varphi_\beta$ with $\beta\geq5$ is possibly non-integrable.

%-------------------------------------%
%-------------- SUBSECTION --------------%
%-------------------------------------%
\subsection{Algebraic entropy}
\label{subsec:AE}
Now we estimate growth of the degree of the iterates of $\varphi_\beta$ to investigate its (non-) integrability.
For this purpose, we prefer the conjugate $\widetilde\varphi_\beta$ to the original $\varphi_\beta$.
Note that a variable transformation does not affect the growth rate of the iterates.

Let us consider the $t$-th iterate
\begin{align*}
\underbrace{\tilde\varphi_\beta\circ\tilde\varphi_\beta\circ\cdots\circ\tilde\varphi_\beta}_{t}:(u^0,v^0)\mapsto(u^t,v^t)
\end{align*}
of the birational map $\widetilde\varphi_\beta$ given by \eqref{eq:bmapbu}.
Introduce the homogeneous coordinate $(u,v)\mapsto[U:V:W]=[u:v:1]$ on $\P^2(\C)$.
The map $\widetilde\varphi_\beta$ then reduces to the following homogeneous one:
\begin{align}
[U^t:V^t:W^t]&\mapsto[U^{t+1}:V^{t+1}:W^{t+1}],
\nn\\
U^{t+1}
&=
V^t\left(W^t\right)^{\beta-4}\left(U^tV^t-\left(W^t\right)^2\right),
\nn\\
V^{t+1}
&=
U^t\left(W^t\right)^{\beta-2}+\left(V^t\right)^{\beta-1},
\label{eq:homogeneousAE}\\
W^{t+1}
&=
\left(W^t\right)^{\beta-3}\left(U^tV^t-\left(W^t\right)^2\right).
\nn
\end{align}
The homogeneous degree of the map $\widetilde\varphi_\beta$ is $\beta-1$.
By applying \eqref{eq:homogeneousAE} repeatedly, $U^t$, $V^t$ and $W^t$ are reduced to the polynomials in the initial values $U^0$, $V^0$ and $W^0$.
By $\delta_\beta^t$, we denote the homogeneous degree of $U^t$, $V^t$ and $W^t$ in $U^0$, $V^0$ and $W^0$.

We then obtain the following proposition.
%//////////////////// THEOREM ////////////////////%
%//////////////////// THEOREM ////////////////////%
%//////////////////// THEOREM ////////////////////%
\begin{proposition}
The homogeneous degree $\delta^t=\delta_\beta^t$ of the birational map $\widetilde\varphi_\beta$ solves the following initial value problem of the second order linear difference equation:
\begin{align}
\begin{cases}
\delta^{t+1}
=
\left(\beta-2\right)\delta^t-\delta^{t-1}
&
(t=1,2,3,\ldots),
\\
\delta^0
=
1,
\\
\delta^1
=
\beta-1.
\end{cases}
\label{eq:lde}
\end{align}
\end{proposition}
%//////////////////// THEOREM ////////////////////%
%//////////////////// THEOREM ////////////////////%
%//////////////////// THEOREM ////////////////////%

%---------- PROOF ----------%
(Proof)\quad
For simplicity, let the values of $U^t$, $V^t$ and $W^t$ be $A$, $B$ and $C$, respectively.
For $t+2$, we compute
\begin{align*}
V^{t+2}
&=
BC^{\beta(\beta-4)+2}
\left(AB-C^2\right)^{\beta-1}
+
\left(AC^{\beta-2}+B^{\beta-1}\right)^{\beta-1}
\\
&=
\sum_{n=0}^{\beta-1}(-1)^{\beta-1-n}\binom{\beta-1}{n}A^nB^{n+1}C^{\beta^2-2\beta-2n}
+
\sum_{n=0}^{\beta-1}\binom{\beta-1}{n}A^nB^{(\beta-1)(\beta-1-n)}C^{(\beta-2)n}
\\
&=
\sum_{n=0}^{\beta-1}\binom{\beta-1}{n}
\left\{
(-1)^{\beta-1-n}C^{\beta(\beta-2)}
+
B^{\beta(\beta-2-n)}C^{n\beta}
\right\}
A^nB^{n+1}C^{-2n}
\\
&=
\left(B^\beta+C^{\beta}\right)
\sum_{n=0}^{\beta-1}
\binom{\beta-1}{n}
\left(
\sum_{m=2}^{\beta-n-1}(-1)^mB^{\beta(\beta-n-1-m)}C^{\beta(m-2)}
\right)
A^{n}B^{n+1}C^{(\beta-2)n},
\end{align*}
where we use the factorization
\begin{align*}
B^{\beta(\beta-n-2)}C^{n\beta}
+
(-1)^{\beta-1-n}C^{\beta(\beta-2)}
&=
C^{n\beta}
\left(B^\beta+C^{\beta}\right)
\sum_{m=2}^{\beta-n-1}(-1)^mB^{\beta(\beta-n-1-m)}C^{\beta(m-2)}.
\end{align*}
We also compute
\begin{align*}
U^{t+2}
&=
\left(
AC^{\beta-2}+B^{\beta-1}
\right)
C^{\beta^2-6\beta+8}\left(AB-C^2\right)^{\beta-3}
\left(
B^{\beta}
+
C^{\beta}
\right),
\\
W^{t+2}
&=
C^{\beta^2-5\beta+5}\left(AB-C^2\right)^{\beta-2}
\left(
B^{\beta}
+
C^{\beta}
\right).
\end{align*}

Remove the common factor $B^{\beta}+C^{\beta}$ from $U^{t+2}$, $V^{t+2}$ and $W^{t+2}$.
It is easy to see that the remaining factors in $U^{t+2}$, $V^{t+2}$ and $W^{t+2}$ are co-prime.
Therefore, we obtain the homogeneous degree of them
\begin{align*}
\delta_\beta^{t+2}
=
\left(\beta^2-3\beta+1\right)\delta_\beta^t
=
\left\{(\beta-2)(\beta-1)-1\right\}\delta_\beta^t
=
(\beta-2)\delta_\beta^{t+1}-\delta_\beta^t,
\end{align*}
where we use $\delta_\beta^{t+1}=(\beta-1)\delta_\beta^t$ (see \eqref{eq:homogeneousAE}).
\qed
%---------- PROOF ----------%

Although we reduce the second order difference equation for $\delta^t_\beta$ in \eqref{eq:lde} from the iterates of $\widetilde\varphi_\beta$, the reason why the recursion relation of $\delta^t_\beta$ turns into such three term relation is not so clear.
It seems that the Laurent phenomenon and the positivity of corresponding cluster algebras lead the result, however, we have not obtained an explanation of such phenomenon based on these properties, yet.

%//////////////////// THEOREM ////////////////////%
%//////////////////// THEOREM ////////////////////%
%//////////////////// THEOREM ////////////////////%
\begin{remark}
In \cite{FH14}, Fordy and Hone applied the algebraic entropy test to nonlinear recurrences for cluster mutation-periodic quivers with period 1 \cite{FM11}, and obtained a series of conjectures implying that four special families of these maps should have zero entropy. 
They examined these families in detail, with many explicit examples given, and showed how they led to discrete integrable systems.
Also, in \cite{HLK19}, Hone, Lampe and Kouloukas investigated the algebraic entropy of coefficient-free cluster algebras, and classified nonlinear recurrences of exchange relations for certain cluster algebras with skew-symmetric exchange matrices.
They reduced piecewise linear recurrence relations for the degrees of the iterates of the seed mutations.
They also computed the algebraic entropy concerning several cluster algebras, explicitly.
\qed
\end{remark}
%//////////////////// THEOREM ////////////////////%
%//////////////////// THEOREM ////////////////////%
%//////////////////// THEOREM ////////////////////%

If we solve the initial value problem \eqref{eq:lde} we obtain the homogeneous degree $\delta_\beta^t$, explicitly.
%//////////////////// THEOREM ////////////////////%
%//////////////////// THEOREM ////////////////////%
%//////////////////// THEOREM ////////////////////%
\begin{proposition}\label{pop:hdegree}
The homogeneous degree $\delta_\beta^t$ of the birational map $\widetilde\varphi_\beta$ is
\begin{align*}
\delta^t_\beta
&=
\begin{cases}
2t+1&\beta=4,\\[5pt]
\DIS\frac{(\nu_++1)(\nu_+)^{t}-(\nu_-+1)(\nu_-)^{t}}{\nu_+-\nu_-}&\beta\geq5,\\
\end{cases}
\end{align*}
where $\nu_\pm$ is the characteristic roots of the difference equation in  \eqref{eq:lde}:
\begin{align}
\nu_\pm=\frac{\beta-2\pm\sqrt{\beta(\beta-4)}}{2}.
\label{eq:characterR}
\end{align}
The sign in the subscript of $\nu_\pm$ corresponds to the one in the right-hand-side.
\end{proposition}
%//////////////////// THEOREM ////////////////////%
%//////////////////// THEOREM ////////////////////%
%//////////////////// THEOREM ////////////////////%

%---------- PROOF ----------%
(Proof)\quad
For $\beta=4$, we have
\begin{align*}
\delta^{t+1}_4-\delta^t_4
&=
\delta^t_4-\delta^{t-1}_4,
\qquad
\delta^1_4=3,
\qquad
\delta^0_4=1
\end{align*}
from \eqref{eq:lde}.
Hence, the sequence $\left\{\delta^t_4\right\}_{t=0,1,\ldots}$ is given by
\begin{align*}
\delta^t_4
=
2t+1.
\end{align*}

Assume $\beta\geq5$.
Then the difference equation in \eqref{eq:lde} reduces to
\begin{align*}
\left(\begin{matrix}
\delta_\beta^{t+1}\\
\delta^t_\beta\\
\end{matrix}\right)
=
\left(\begin{matrix}
\beta-2&-1\\
1&0\\
\end{matrix}\right)
\left(\begin{matrix}
\delta_\beta^{t}\\
\delta_\beta^{t-1}\\
\end{matrix}\right)
=
\left(\begin{matrix}
\beta-2&-1\\
1&0\\
\end{matrix}\right)^t
\left(\begin{matrix}
\beta-1\\
1\\
\end{matrix}\right).
\end{align*}
Now we put
\begin{align*}
M
:=
\left(\begin{matrix}
\beta-2&-1\\
1&0\\
\end{matrix}\right).
\end{align*}
The eigenvalues of $M$ are $\nu_\pm$ and the corresponding eigenvectors are
\begin{align*}
\bp_+
=
\left(\begin{matrix}
\nu_+\\
1\\
\end{matrix}\right)
\qquad
\mbox{and}
\qquad
\bp_-
=
\left(\begin{matrix}
\nu_-\\
1\\
\end{matrix}\right),
\end{align*}
respectively.
By using the regular matrix $P:=\left(\bp_+\bp_-\right)$, we have
\begin{align*}
\left(\begin{matrix}
\delta_\beta^{t+1}\\
\delta^t_\beta\\
\end{matrix}\right)
&=
P
\left(\begin{matrix}
(\nu_+)^t&0\\
0&(\nu_-)^t\\
\end{matrix}\right)
P^{-1}
\left(\begin{matrix}
\beta-1\\
1\\
\end{matrix}\right)
\\
&=
\frac{1}{\nu_+-\nu_-}
\left(\begin{matrix}
(\nu_+)^{t+1}-(\nu_-)^{t+1}&-(\nu_+)^{t}+(\nu_-)^{t}\\
(\nu_+)^{t}-(\nu_-)^{t}&-(\nu_+)^{t-1}+(\nu_-)^{t-1}\\
\end{matrix}\right)
\left(\begin{matrix}
\beta-1\\
1\\
\end{matrix}\right).
\end{align*}
Therefore, we obtain
\begin{align*}
\delta^t_\beta
&=
\frac{(\nu_++1)(\nu_+)^{t}-(\nu_-+1)(\nu_-)^{t}}{\nu_+-\nu_-},
\end{align*}
where we use $\nu_+\nu_-=1$.
\qed
%---------- PROOF ----------%

%//////////////////// THEOREM ////////////////////%
%//////////////////// THEOREM ////////////////////%
%//////////////////// THEOREM ////////////////////%
\begin{remark}
From the characteristic roots $\nu_\pm$ (see \eqref{eq:characterR}), we compute
\begin{align*}
\nu_\pm+1
=
\frac{\beta\pm\sqrt{\beta(\beta-4)}}{2},
\qquad
\nu_+-\nu_-
=
\sqrt{\beta(\beta-4)}
\end{align*}
and
\begin{align*}
(\nu_+)^{t}-(\nu_-)^{t}
&=
\frac{1}{2^{t-1}}
\sum_{m=0}^{\lfloor t-1/2\rfloor}\binom{t}{2m+1}(\beta-2)^{t-2m-1}\sqrt{\beta(\beta-4)}^{2m+1},
\\
(\nu_+)^{t}+(\nu_-)^{t}
&=
\frac{1}{2^{t-1}}
\sum_{m=0}^{\lfloor t/2\rfloor}\binom{t}{2m}(\beta-2)^{t-2m}\sqrt{\beta(\beta-4)}^{2m}.
\end{align*}
We then obtain
\begin{align}
\delta_\beta^t
&=
\left(\frac{\beta-2}{2}\right)^t
\left\{
\frac{\beta}{\beta-2}
\sum_{m=0}^{\lfloor t-1/2\rfloor}\binom{t}{2m+1}\left(\frac{\beta(\beta-4)}{(\beta-2)^2}\right)^m
+
\sum_{m=0}^{\lfloor t/2\rfloor}\binom{t}{2m}\left(\frac{\beta(\beta-4)}{(\beta-2)^2}\right)^m
\right\}.
\label{eq:deltaexplicit}
\end{align}
Thus the square roots in $\delta_\beta^t$ are removed.
\qed
\end{remark}
%//////////////////// THEOREM ////////////////////%
%//////////////////// THEOREM ////////////////////%
%//////////////////// THEOREM ////////////////////%

Define the algebraic entropy $\mathfrak{E}_\beta$ \cite{BV99} of the birational map $\widetilde\varphi_\beta$ to be
\begin{align*}
\mathfrak{E}_\beta
:=
\lim_{t\to\infty}\frac{1}{t}\log \delta_\beta^t.
\end{align*}
We immediately obtain the explicit form of $\mathfrak{E}_\beta$ from proposition \ref{pop:hdegree}.
%//////////////////// THEOREM ////////////////////%
%//////////////////// THEOREM ////////////////////%
%//////////////////// THEOREM ////////////////////%
\begin{proposition}
The algebraic entropy $\mathfrak{E}_\beta$ of the birational map $\widetilde\varphi_\beta$ is given by
\begin{align}
\mathfrak{E}_\beta
&=
\begin{cases}
0&\beta=4,\\
\log\nu_+&\beta\geq5.\\
\end{cases}
\label{eq:AEexplicit}
\end{align}
\end{proposition}
%//////////////////// THEOREM ////////////////////%
%//////////////////// THEOREM ////////////////////%
%//////////////////// THEOREM ////////////////////%

%---------- PROOF ----------%
(Proof)\quad
For $\beta=4$, we have $\delta_4^t=2t+1$ by proposition \ref{pop:hdegree}.
It immediately follows $\mathfrak{E}_\beta=0$.

Assume $\beta\geq5$.
By proposition \ref{pop:hdegree},  we compute
\begin{align*}
\log \delta_\beta^t
&=
\log\frac{(\nu_++1)(\nu_+)^{t}-(\nu_-+1)(\nu_-)^{t}}{\nu_+-\nu_-}
\\
&=
\log\left\{(\nu_++1)e^{t\log\nu_+}-(\nu_-+1)e^{t\log\nu_-}\right\}
-
\frac{1}{2}\log\left(\beta(\beta-4)\right).
\end{align*}
Since $\beta\geq5$, we have $0<\nu_-<1<\nu_+$, and hence have $\log\nu_-<0<\log\nu_+$.
Therefore, we obtain
\begin{align*}
\log\left\{(\nu_++1)e^{t\log\nu_+}-(\nu_-+1)e^{t\log\nu_-}\right\}
&<
\log\left\{(\nu_++1)e^{t\log\nu_+}\right\}
\\
&=
t\log\nu_+
+
\log(\nu_++1)
\end{align*}
and
\begin{align*}
\log\left\{(\nu_++1)e^{t\log\nu_+}-(\nu_-+1)e^{t\log\nu_-}\right\}
&>
\log\left\{(\nu_++1)e^{t\log\nu_+}-2\right\}
\\
&>
\log e^{t\log\nu_+}
\\
&=
t\log\nu_+
\end{align*}
for sufficiently large $t$.
Thus we conclude that 
\begin{align*}
\mathfrak{E}_\beta
=
\lim_{t\to\infty}\frac{1}{t}\log \delta_\beta^t
=
\log\nu_+
=
\log\frac{\beta+\sqrt{\beta(\beta-4)}}{2}
\end{align*}
holds.
The algebraic entropy $\mathfrak{E}_\beta$ is nothing but the logarithm of the spectral radius of the matrix $M$ given by the linear difference equation \eqref{eq:lde} (see the proof of proposition \ref{pop:hdegree}).
\qed
%---------- PROOF ----------%

Thus we show that the birational map $\widetilde\varphi_\beta$ with $\beta\geq5$ has the positive algebraic entropy $\mathfrak{E}_\beta=\log\nu_+$; whereas $\mathfrak{E}_\beta=0$ when $\beta\leq4$, which is consistent with the fact that the map $\widetilde\varphi_\beta$ with $\beta\leq4$ generates integrable systems.

%//////////////////// THEOREM ////////////////////%
%//////////////////// THEOREM ////////////////////%
%//////////////////// THEOREM ////////////////////%
\begin{example}
For $\beta=5$, we have
\begin{align*}
\delta^t_5
&=
\left(\frac{3}{2}\right)^t
\left\{
\frac{5}{3}
\sum_{m=0}^{\lfloor t-1/2\rfloor}\binom{t}{2m+1}\left(\frac{5}{9}\right)^m
+
\sum_{m=0}^{\lfloor t/2\rfloor}\binom{t}{2m}\left(\frac{5}{9}\right)^m\right\}
\end{align*}
(see \eqref{eq:deltaexplicit}).
It immediately follows
\begin{align*}
\left\{\delta^t_5\right\}_{t=0,1,2,\ldots}
=
\left\{1, 4, 11, 29, 76, 199, 521, 1364, 3571, 9349, 24476,\ldots\right\}.
\end{align*}
Moreover, we have
\begin{align*}
\mathfrak{E}_5
=
\log\frac{3+\sqrt{5}}{2}
=
0.962424\ldots.
\end{align*}
Thus the algebraic entropy is positive.
\qed
\end{example}
%//////////////////// THEOREM ////////////////////%
%//////////////////// THEOREM ////////////////////%
%//////////////////// THEOREM ////////////////////%

%-------------------------------------%
%-------------- SUBSECTION --------------%
%-------------------------------------%
\subsection{Families of integrable and non-integrable systems}
\label{subsec:CCA}
We establish two remarkable properties of the difference equation \eqref{eq:2dds} with $\beta\geq5$ in the previous subsections.
One is that the birational map $\widetilde\varphi_\beta$ equivalent to \eqref{eq:2dds} has the following positive algebraic entropy
\begin{align}
\mathfrak{E}_\beta=\log\frac{\beta+\sqrt{\beta(\beta-4)}}{2};
\label{eq:AEBeta}
\end{align}
and the other is that the birational map $\varphi_\beta$ conjugate to $\widetilde\varphi_\beta$ fails the singularity confinement test.
These properties enables us to conclude that the equation \eqref{eq:2dds} with $\beta\geq5$ is non-integrable, whereas it is integrable when $\beta\leq4$.

The property of the one-parameter family of the equation \eqref{eq:2dds}, whose dynamics varies from integrable to non-integrable depending on the parameter $\beta$, is similar to the extended Hietarinta-Viallet equation introduced by Kanki, Mase and Tokihiro:
\begin{align}
x_{n-1}+x_{n+1} =x_n + \frac{1}{x_n^k}
\qquad
(k\in\N).
\label{eq:EHV}
\end{align}
The second-order difference equation \eqref{eq:EHV} with $k=1$ is integrable.
(Remark that \eqref{eq:EHV} with $k=1$ passes the singularity confinement test.)
With odd $k\geq3$, the difference equation \eqref{eq:EHV} fails the singularity confinement test and the algebraic entropy is a positive number
\begin{align}
\lambda_k:=\log\frac{k+\sqrt{k(k+4)}}{2}.
\label{eq:AEk}
\end{align}
Thus the dynamical property of \eqref{eq:EHV} with odd $k$ coincides with the one of our difference equation \eqref{eq:2dds}.
It should be noted that the algebraic entropies \eqref{eq:AEBeta} and \eqref{eq:AEk} are in the following relation
\begin{align*}
e^{\mathfrak{E}_\beta}
=
e^{\lambda_{\beta-4}}+2
\end{align*}
for odd $\beta\geq7$.

On the other hand, the difference equation \eqref{eq:EHV} with even $k\geq2$ passes the singularity confinement test although the algebraic entropy is still a positive number
\begin{align*}
\mu_k:=\log\frac{k+1+\sqrt{(k-1)(k+3)}}{2}.
\end{align*}
Due to this property, the difference equation \eqref{eq:EHV} is referred to as the extended ``Hietarinta-Viallet equation'' \cite{HV98}.
Remark that we also have
\begin{align*}
e^{\mathfrak{E}_\beta}
=
e^{\mu_{\beta-3}}+1
\end{align*}
for odd $\beta\geq5$.

\begin{figure}[htb]
\centering
\includegraphics[scale=1,pagebox=cropbox]{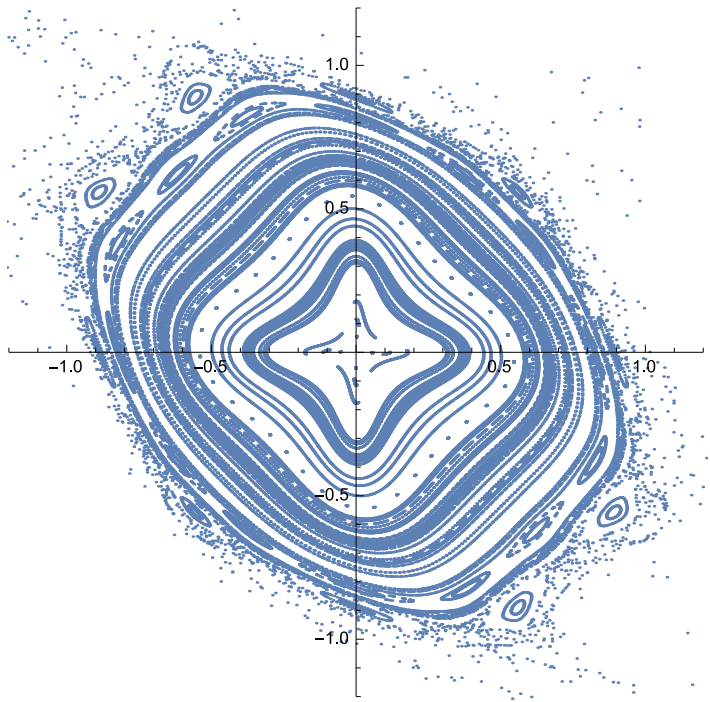}
\caption{The orbits of the 1000 iterates of $\widetilde\varphi_6$ ($\beta=6$) from randomly chosen 100 initial values.}
\label{fig:invcurve}
\end{figure}

Finally, we observe that the birational map $\widetilde\varphi_\beta$ with $\beta\geq5$ exhibits numerical chaos.
It can be seen when we draw a picture of some orbits of the birational map $\widetilde\varphi_\beta$, see  figure \ref{fig:invcurve}.
This figure is obtained with $\beta= 6$. 
The picture is characteristic of chaotic behavior of a two dimensional system.

%//////////////////// FIGURE ////////////////////%
%//////////////////// FIGURE ////////////////////%

%-------------------------------------%
%-------------- SECTION --------------%
%-------------------------------------%
\section{Concluding remarks}
\label{sec:CR}
We study the one-parameter family of second order nonlinear difference equations \eqref{eq:2dds} arising from the seed mutations of cluster algebras imposing the initial seed \eqref{eq:initialseed}.
The equation \eqref{eq:2dds} with $\beta=1$, $2$ or $3$ has the solution of period $5$, $3$ or $4$ for any initial value, respectively.
The equation \eqref{eq:2dds} with $\beta\geq4$, however, is periodic no longer; nevertheless, we can show its integrability only when $\beta=4$ by using the conserved quantity.
Simultaneously, we show that the equation \eqref{eq:2dds} with $\beta\geq4$ has two remarkable properties.
One is that it never passes the singularity confinement test, a criterion for possibility to construct the space of the initial values.
This fact is shown by solving the system of linear difference equation \eqref{eq:recdegree} for the minimal degrees of the parameter $\varepsilon$ in the homogeneous variables of the birational map $\varphi_\beta$ equivalen to \eqref{eq:2dds}.
Failure of the singularity confinement is consistent with the linearizability of $\varphi_\beta$ with $\beta=4$.
The other remarkable property is that we can explicitly compute the algebraic entropy of $\varphi_\beta$ by using the initial value problem \eqref{eq:lde} of the second order linear difference equation. 
We obtain the algebraic entropy given by the logarithm \eqref{eq:AEexplicit} of the characteristic root of the linear difference equation in \eqref{eq:lde}.
The algebraic entropy with $\beta=4$ is zero, which suggests integrability of \eqref{eq:2dds}, too; whereas, the algebraic entropy with $\beta\geq5$ is positive, which suggests non-integrability of \eqref{eq:2dds}.
We also observe that the orbits of $\varphi_\beta$ with $\beta\geq5$ exhibit numerical chaos.
Thus the one-parameter family of the difference equation \eqref{eq:2dds} possesses both integrable ($\beta\leq4$) and non-integrable ($\beta\geq5$) equations.

Although relations between integrable systems and cluster algebras have intensively studied for a dozen years, as far as the authors know, non-integrable systems concerning cluster algebras have not been studied so much.
Nevertheless, such non-integrable systems are more universal than integrable ones; the difference equation \eqref{eq:2dds} is non-integrable for infinitely many $\beta\in\N$ except four cases ($\beta=1,2,3,4$) (see also \cite{HLK19}).
Moreover, as we have shown in this paper, non-integrable systems concerning cluster algebras are expected to have fine properties such as computability of the algebraic entropy.
We plan to study non-integrable systems concerning cluster algebras more precisely and to clarify roles of the Laurent phenomenon and positivity in discrete dynamical systems.

\end{document}